\documentclass[10pt, twocolumn, aps, prx, superscriptaddress]{revtex4-2}
\usepackage[dvipsnames]{xcolor}
\usepackage{graphicx}
\usepackage{amsmath, amssymb}
\usepackage{braket, upgreek}
\usepackage{BOONDOX-cal}
\usepackage{enumitem}
\setlist[enumerate]{left=1mm}
\usepackage[unicode=true, pdfusetitle,
            bookmarks=true, bookmarksnumbered=false,
            bookmarksopen=true, breaklinks=false,
            bookmarksopenlevel=1,
            pdfborder={0 0 1}, backref=false, colorlinks=true]{hyperref}
\hypersetup{pdfborderstyle=, linkcolor=blue, 
            urlcolor=blue, citecolor=blue}
\usepackage{etoolbox}

\usepackage[normalem]{ulem} 
\newif\ifshowchanges
\showchangesfalse  
\newcommand{\changecolor}{BrickRed}

\newcommand{\tagbubble}[1]{}

\newcommand{\add}[2][]{%
\ifshowchanges
{\textcolor{\changecolor}{#2}\if\relax\detokenize{#1}\relax\else\tagbubble{#1}\fi}%
\else
#2%
\fi
}
\newcommand{\del}[2][]{%
\ifshowchanges
{\textcolor{BrickRed}{\sout{#2}}\if\relax\detokenize{#1}\relax\else\tagbubble{#1}\fi}%
\else
\fi
}
\newcommand{\rep}[3][]{%
\ifshowchanges
{\textcolor{\changecolor}{\sout{#2}}\,\textcolor{\changecolor}{#3}\if\relax\detokenize{#1}\relax\else\tagbubble{#1}\fi}%
\else
#3%
\fi
}

\newcommand{\e}{\mathcal{e}}
\newcommand{\p}{\mathcal{p}}
\newcommand{\ep}{\mathcal{e}{\text -}\mathcal{p}}
\newcommand{\pe}{\mathcal{p}{\text -}\mathcal{e}}

\date{\today}

\begin{document}

\title{Noise-enhanced Ballistic Expansion of Polariton Wave-packets in a Multimode Cavity}
\author{Ilia Tutunnikov}
\thanks{These authors contributed equally to this work.}
\affiliation{Department of Chemistry, Massachusetts Institute of Technology, Cambridge,
Massachusetts 02139, USA}
\affiliation{ITAMP, Center for Astrophysics \textbar{} Harvard \& Smithsonian,
Cambridge, Massachusetts 02138, USA}

\author{Md Qutubuddin}
\thanks{These authors contributed equally to this work.}
\affiliation{Beijing Computational Science Research Center, Beijing 100193, China}

\author{H. R. Sadeghpour}
\affiliation{ITAMP, Center for Astrophysics \textbar{} Harvard \& Smithsonian,
Cambridge, Massachusetts 02138, USA}

\author{Jianshu Cao}
\email{jianshu@mit.edu}
\affiliation{Department of Chemistry, Massachusetts Institute of Technology, Cambridge,
Massachusetts 02139, USA}

\begin{abstract}
Advances in optical measurements enable precise tracking of cavity polariton wave-packets across broad spatial and temporal ranges, but how dephasing reshapes their real-space dynamics over multiple time scales remains unclear.  Here we show, using a stochastic multimode Tavis–Cummings model, that dephasing noise leads to a robust hierarchy of dynamical regimes comprising Rabi oscillation damping, center-of-mass slowdown, population relaxation, and ballistic-to-diffusive crossover, in the order of increasing time scales. We further predict that dephasing can enhance ballistic spreading and sustain it far beyond the microscopic dephasing time by two orders of magnitude. These predictions agree with recent microscopy measurements and provide experimentally testable guidance for engineering energy transport in polaritonic platforms.
\end{abstract}
\maketitle

\section{Introduction}

Strong light-matter interactions are promising for controlling and enhancing
material properties and engineering emerging devices. Cavities are especially 
effective for achieving a strong collective coupling between material 
excitations and cavity photons, giving rise to hybrid excitations termed 
cavity polaritons.
Polaritons exhibit diverse phenomena with important applications 
across photonics and materials science~\cite{nature_materialreview, 
Garcia2021, Sanvitto2016}. Over the past decade, recent experiments in cavity-based 
platforms have shown modifications to chemical dynamics and enhanced energy 
transfer~\cite{Son2022, DelPo2021, Lerario2017, Zhong2016, Coles2014, saez2019, Du2018, Reitz2018, jianshu2022jpcl},
exciton~\cite{Khazanov2023, tichauer2023, Hou2020, Rozenman2018, Wenus2006}
and charge transport \cite{Liu2022, Krainova2020, Nagarajan2020, Orgiu2015}.
These experimental advancements have led to the emergence of polaritonic chemistry 
as a rapidly growing research area~\cite{LiTao2022, Ribeiro2018, Herrera2016}. 
To complement the experimental breakthroughs, theoretical models have been 
developed to account for the details of the emitters and complex environments~\cite{Tutunnikov2023, Amir2009}, 
allowing them to describe the subtleties of exciton-polariton systems~\cite{Alvaro_2024, 
AroeiraRibeiro2023, Arnardottir2020, Hagenmuller2017, 
Gonzalez-Ballestero2015, Feist2015, schachenmayer2015,Engelhardt2022, Engelhardt2023, Wu2024}.

Modern microscopy techniques allow tracking of the evolution of exciton-polariton 
wave packets with high spatiotemporal resolution~\cite{Balasubrahmaniyam2023, 
Jin2023, Xu2023, Pandya2022}. 
These experiments motivate the theoretical analysis of polariton-mediated transport 
in multi-mode cavities. While single-mode models, like Dicke~\cite{dicke54} and 
Tavis-Cummings (TC)~\cite{tavis1968exact}, allow for simplified analyses, they 
predict that localized excitations can instantaneously travel to distant emitters, 
which is unphysical. 
In contrast, multi-mode models with physical photonic dispersion satisfy the relativistic constraints~\cite{Engelhardt2022, Engelhardt2023}.

A comprehensive and rigorous treatment of polariton dynamics in multi-mode cavities is 
challenging due to the typically large number of emitters and cavity modes. 
Moreover, the complexities of realistic systems, including cavity losses, 
disorder, and environmental dephasing~\cite{Moix2013, tutunnikov2024unusualdiffusivity},
make accurate theoretical description even more demanding.
The effects of static and dynamic disorder in polariton systems have been 
explored theoretically using various approaches, 
including perturbation theory \cite{Agranovich2003, Litinskaya2004}, exact 
diagonalization and integration \cite{AroeiraRibeiro2023, Engelhardt2023}, 
mean-field-based approaches \cite{Strashko2018, Sokolovskii2023, osipov2023}, 
and ab initio simulation \cite{Alvertis2020, schafer2019}.
Despite these advancements, a complete dynamic picture covering all the stages of
evolution is still lacking. This includes a detailed description of the transition 
between the ballistic and diffusive regimes of the wave packet expansion.

Here, we provide a complete analysis of the spatiotemporal
dynamics of polariton wave packets in multi-mode cavities. We adopt 
the stochastic multi-mode TC model to incorporate both the continuum
of photon modes and the stochastic dephasing noise acting on the emitters. 
This stochastic model is appropriate when the emitters are coupled to 
a thermal bath at a relatively high temperature.
To make the analysis feasible, we study an effectively one-dimensional (1D) microcavity with an embedded 1D lattice of identical emitters~\cite{Engelhardt2023}.
Although the emitters do not interact directly, their coupling to the 
cavity modes enables the propagation of excitonic wave packets.

The developed model combines methodological advances with
experimentally relevant predictions. Our quantum master equation enables
efficient large-scale simulations of multimode cavity systems with emitter
dephasing and provides a practical framework for extensions that include cavity
loss, static disorder, and direct inter-emitter couplings. We find that noisy
dynamics follow a robust hierarchy of relaxation stages, from underdamped Rabi
oscillations to center-of-mass slowdown, population relaxation, and
ballistic-to-diffusive crossover. The analysis further shows that ballistic
spreading can persist for times far longer than the microscopic dephasing time
$\Gamma^{-1}$, while the spreading rate is enhanced and the long-time diffusion
constant is reduced relative to the noise-free case.
\begin{figure}
\begin{centering}
\includegraphics{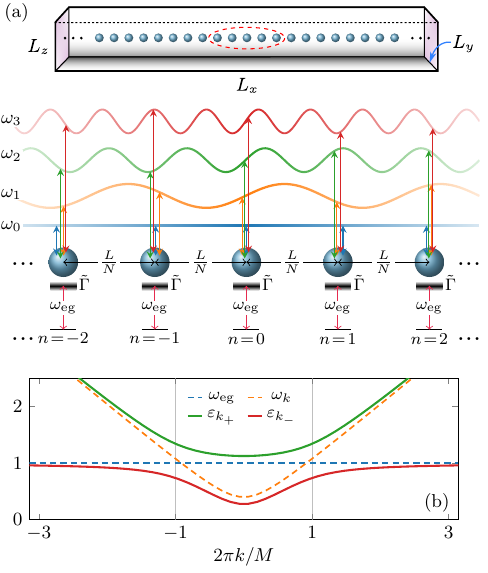} 
\end{centering}
\caption{\textbf{Multimode cavity model and polariton
dispersion.} (a)\;Schematic illustration of the multimode TC
model. $N$ identical two-level-system (TLS) emitters (blue spheres) are
arranged in a 1D lattice along the $x$ axis of an effectively 1D optical
cavity of length $L \equiv L_x$ with strong transverse confinement
($L_y, L_z \ll L_x$). The light-blue background denotes the cavity volume, and
the pale red vertical bands indicate representative stochastic fluctuations of
the emitter transition energies. The lower inset shows five example emitters
near the cavity center. Their transition energy is $\hbar\omega_{\mathrm{eg}}$,
the lattice spacing is $L/N$, and only couplings to the example cavity modes
with energies $\hbar\omega_0,\dots,\hbar\omega_3$ are drawn for clarity.
Within the Haken--Strobl--Reineker model discussed in the Results subsection on
stochastic noise acting on emitters, the transition energy fluctuates about the
average value $\omega_{\mathrm{eg}}$ with autocorrelation function
$\tilde{\Gamma}\delta(t)$. (b)\;Illustrative single-excitation dispersion. The
dashed lines show the bare emitter energy $\omega_{\mathrm{eg}}$ and the photon
dispersion $\omega_k$, while the solid curves show the upper and lower
polariton branches $\varepsilon_{k_\pm}$ as functions of $2\pi k/M$ (example
parameters $\omega_{\mathrm{eg}}=1$, $q_c=0.4$, $g=0.3$).}
\label{fig:FIG1}
\end{figure}

\section{Methods \label{sec:System-definition}}

We consider a 1D lattice of $N\gg1$ identical, non-interacting emitters 
(two-level systems, TLS) with transition energy $\hbar\omega_{\mathrm{eg}}$ 
embedded in effectively 1D cavity of length $L \equiv L_x$, i.e., 
multi-mode TC model (Fig.~\ref{fig:FIG1};\,(a) schematic of the cavity and emitters coupling to modes; (b) illustrative single-excitation dispersion of $\omega_k$, $\omega_{\mathrm{eg}}$, and $\varepsilon_{k_\pm}$).
 Cavity mode frequencies 
are given by 
\begin{equation}
\tilde{\omega}_{k}=\sqrt{c^{2}\left(\frac{2\pi k}{L}\right)^{2}+\tilde{\omega}_{c}^{2}},
\label{eq:modes-energies}
\end{equation}
where $k$ is an integer, $c$ is the speed of light, and $\tilde{\omega}_{c}>0$
arises due to the spatial confinement of the field in the $y$ and
$z$ directions. When $N=M$ (i.e., $\uprho=1$) and $k$ spans $-M/2\le k\le M/2$, the photon energies range from $\tilde{\omega}^{min}_{k} = 0.4 \omega_{eg}$ up to $\tilde{\omega}^{\max}_{k}= 3.16 \omega_{eg}$. Assuming the emitters are located at positions $r_{n}=nL/N$
($n$ is an integer, $-N/2\leq n\leq N/2$), the system's Hamiltonian is given
by \cite{Engelhardt2023, AroeiraRibeiro2023, AroeiraRibeiro2024}
\begin{align}
\hat{\tilde{H}} & =\sum_{k}\hbar\tilde{\omega}_{k}\ket{a_{k}}\bra{a_{k}}+\sum_{n=-N/2}^{N/2}\hbar\omega_{\mathrm{eg}}\ket{b_{n}}\bra{b_{n}}\nonumber \\
 & +\frac{1}{\sqrt{L}}\sum_{k,n} \tilde{g}_k e^{i\frac{2\pi k}{L}r_{n}}\ket{a_{k}}\bra{b_{n}}+\mathrm{c.c}.,
\label{eq:defaul-Hamiltonian}
\end{align}
where $\ket{a_{k}}$ is the cavity state with a single excitation
in photon mode $k$, $\ket{b_{n}}$ is the matter subsystem state with a 
single excited emitter at site $n$, and $\tilde{g}>0$ is the matter-field
coupling constant. For analytical tractability, we assumed a uniform coupling strength 
\( \tilde{g}_k = \tilde{g}\). Our analysis is restricted to the single excitation manifold. 
In general, the light--matter coupling can be mode- and gauge-dependent (e.g., in the Power--Zienau--Woolley gauge $g_k\propto\sqrt{\omega_k}$). Here we treat $\tilde g_k$ as an effective constant for the near-resonant modes that dominate the wave packets considered, which keeps the model analytically and numerically tractable; allowing a mild $k$-dependence would primarily renormalize the polariton composition and quantitative timescales rather than the qualitative sequence of dynamical stages discussed below.

Hereafter, energy, length, frequency, and time are measured
in units of $\hbar\omega_{\mathrm{eg}}$, $\ell \equiv c/\omega_{\mathrm{eg}}$,
$\omega_{\mathrm{eg}}$, and $1/\omega_{\mathrm{eg}}$ respectively. 
$\hbar$ is set to unity. 
Letting $L=M\ell$, the dimensionless mode frequency is given by
\begin{equation}
\omega_{k}=\sqrt{\left(\frac{2\pi k}{M}\right)^{2}+\omega_c^2},
\label{eq:omega_k-dimensionless}
\end{equation}
where $\omega_{c}=\tilde{\omega}_{c}/\omega_{\mathrm{eg}}$. $\hat{\tilde{H}}$ can be simplified using the Bloch states as the basis for
the emitters' subsystem. The transformed dimensionless Hamiltonian
is given by \add{(see Supplementary Note 1 for
details)} 
\begin{align}
\hat{H} 
& =
\sum_{k=-N/2}^{N/2}\left(\omega_{k}\ket{a_{k}}\bra{a_{k}}+\ket{b_{k}}\bra{b_{k}}\right)
\nonumber \\
&+g\sum_{k=-N/2}^{N/2}\left(\ket{a_{k}}\bra{b_{k}}+\ket{b_{k}}\bra{a_{k}}\right),
 \label{eq:H-transformed}
\end{align}
where $g=\tilde{g}\sqrt{\uprho}/(\hbar\omega_{\mathrm{eg}}
\sqrt{\ell})=\tilde{g}\sqrt{N/M}/(\hbar\omega_{\mathrm{eg}}\sqrt{\ell})$ 
is the dimensionless collective coupling constant,
and $\uprho=N/M$ is the dimensionless number density.
This form of $\hat{H}$ reveals that the system can be viewed
as a collection of TLSs, each describing the interaction between the
$k$-th exciton and the $k$-th cavity mode. 
Throughout the paper, we set the number of cavity modes equal to the number
of emitters, such that $-N/2 \leq k \leq N/2$.
We focus on $\uprho=1$ (equal numbers of emitters and cavity modes), for which the coupled exciton--photon mode pairs span the single-excitation manifold considered here. For $\uprho>1$ (more emitters than cavity modes), an additional manifold of excitonic dark states appears; in the absence of noise these states do not couple to photons, but a generic localized excitation may still have overlap with them, reducing the propagating bright-state weight. In the presence of dephasing or vibronic couplings, bright--dark mixing can further modify transport.

The TLS eigenstates are the upper ($+$) and lower ($-$) polaritons 
\begin{subequations}
\begin{align}
\ket{v_{k_{\pm}}} & =\begin{cases}
-\sin(\theta_{k})\ket{a_{k}}+\cos(\theta_{k})\ket{b_{k}}, & +\\
\cos(\theta_{k})\ket{a_{k}}+\sin(\theta_{k})\ket{b_{k}}, & -
\end{cases},\\
\theta_{k} & =\frac{1}{2}\arctan\left[\frac{2g}{\omega_{k}-\omega_{\mathrm{eg}}}\right]+\frac{\pi}{2}\Theta(\omega_{k}-\omega_{\mathrm{eg}}),
\end{align}
\end{subequations}
where $\Theta(u)=0$ when $u<0$, and $\Theta(u)=1$ when $u>0$.
The TLS eigenenergies, i.e., the polaritonic branches, are given by
\begin{subequations}
\label{eq:Polaritons-def} 
\begin{align}
\varepsilon_{k_{\pm}} & =\frac{\omega_{k}+\omega_{\mathrm{eg}}}{2}\pm\frac{\Delta_{k}}{2},\label{eq:UL-polariton-energies}\\
\Delta_{k} & \equiv\varepsilon_{k_{+}}-\varepsilon_{k_{-}}=\sqrt{4g^{2}+(\omega_{k}-\omega_{\mathrm{eg}})^{2}},\label{eq:VRS}
\end{align}
\end{subequations}
where $\Delta_{k}$ is the vacuum Rabi splitting (VRS). The polariton spectrum consists of two branches, $\varepsilon_{k_\pm}$, separated by the vacuum Rabi splitting $\Delta_k$. The corresponding group velocities,
$v_\pm(k)=\partial_k\varepsilon_{k_\pm}$, govern the ballistic propagation of polariton wave packets analyzed below.
An illustrative plot of $\omega_k$, $\omega_{\mathrm{eg}}$, and $\varepsilon_{k_\pm}$ vs.\ $2\pi k/M$ is shown in Fig.~\ref{fig:FIG1}(b).
A detailed discussion of the dispersion relations, vacuum Rabi splitting,
and group-velocity structure for the present model and parameters is given
in the corresponding noise-free analysis.

\subsection{Definition of relevant quantities \label{sec:Dynamics-Isolated}}

We analyze the dynamics generated by the Hamiltonian
$\hat H$ in Eq.~\eqref{eq:H-transformed} in the absence of noise.
The purpose of this section is to establish a rigorous reference
description of coherent polariton-mediated dynamics, formulated in
terms of the moments of the population and the total excitonic
population.
These excitonic quantities (here, emitters refer to the $N$ molecules, and excitons to the excited molecules) will serve as benchmarks for the noise-induced effects discussed in subsequent sections.

We consider a single excitation initially prepared in the emitter
subsystem, while all cavity modes are unoccupied.
The matter excitation is initialized as a traveling Gaussian wave
packet in real space.
In momentum space, the corresponding probability distribution reads
\begin{equation}
P_{k}(0)=\frac{w}{\sqrt{\pi}}e^{-w^{2}(k-p)^{2}},
\label{eq:P_k(0)}
\end{equation}
where $w$ is the dimensionless real-space width and $p$ is the mean
momentum.

To connect the excitonic population defined above with real-space dynamics,
we introduce the corresponding initial polariton wave packet.
In the single-excitation manifold, the state can be written in the polariton
eigenbasis as
\begin{equation}
\psi(t) = \sum_k \psi_k(0)\, e^{-i\varepsilon_{k,\pm} t}\ket{v_{k,\pm}},
\label{eq:polariton-wave-packet}
\end{equation}
where $\psi_k(0)=\sqrt{\tfrac{w}{\sqrt{\pi}}}\,
e^{-\tfrac{w^2}{2}(k-p)^2}$ is the initial Gaussian distribution in momentum
space, $w$ is the real-space width, and $p$ is the mean momentum.
Here, $\ket{v_{k,\pm}}$ denote the upper ($+$) and lower ($-$) polariton
eigenstates.

Projecting $\psi(t)$ onto the excitonic subspace yields two contributions
associated with the upper and lower polariton branches, weighted by the
corresponding mixing coefficients.
For a narrow wave packet ($w\gg 1$), these weights may be evaluated at the mean
momentum $p$, leading to two wave packets propagating with group velocities
$v_\pm=\partial_k\varepsilon_{k,\pm}|_{k=p}$.

As a result, the real-space excitonic probability density is well approximated
by a superposition of two Gaussian wave packets, each associated with one
polariton branch, whose centers move ballistically while their widths increase
in time due to dispersion.
\subsection{Moments of the exciton population distribution}

To characterize the spatiotemporal evolution of the matter distribution,
we focus on the first two moments of the real-space exciton population distribution.
In the continuum limit, the diagonal element of the real-space excitonic
density matrix can be written in terms of the momentum-space density
matrix as
\begin{align}
\rho_{x,x}(t)
&=\frac{1}{2\pi}\int e^{-ixk}\rho^{\e}_{k,k'}(t)e^{ixk'}\,dk\,dk' \nonumber\\
&=\frac{1}{2\pi}\int e^{ixq}\rho^{\e}_{k,k+q}(t)\,dk\,dq .
\end{align}
Using the identities
$\int x e^{ixq}dx = i\,\delta'(q)$ and
$\int x^2 e^{ixq}dx = -\delta''(q)$,
the first and second moments of the emitter population are obtained as
\begin{align}
\braket{n(t)}
&=\int x\,\rho_{x,x}(t)\,dx
= -i\int\left.\frac{\partial\rho^{\e}_{k,k+q}}{\partial q}\right|_{q=0}dk,
\label{eq:<n>-kk-defintion}\\
\braket{n^{2}(t)}
&=\int x^{2}\rho_{x,x}(t)\,dx
= -\int\left.\frac{\partial^{2}\rho^{\e}_{k,k+q}}{\partial q^{2}}\right|_{q=0}dk .
\label{eq:<n^2>-kk-definition}
\end{align}

The spatial spreading of the matter wave packet is quantified by the
(exciton-normalized) width
\begin{equation}
W(t)=\sqrt{
\frac{\braket{n^{2}(t)}}{P_{ex}(t)}
-\frac{\braket{n(t)}^{2}}{P_{ex}^{2}(t)}
-\frac{w^{2}}{2}},
\label{eq:W}
\end{equation}
where $P_{ex}(t)$ denotes the total population in the emitter subsystem.
\\
\section{Results}
\subsection{Dynamics in presence of stochastic noise acting on emitters \label{sec:Dynamics-Noise}}

This section analyzes the probability distribution of the emitters within
a stochastic multi-mode Tavis--Cummings (TC) model. We adopt the
Haken--Strobl--Reineker (HSR) approach to describe noise
\cite{KenkreReineker1982, Lee2015}, where each emitter's transition energy
acquires an independent stochastic contribution $ \xi_{n}(t) $. This noise
is taken to be white/uncorrelated, with an auto-correlation function
$ \overline{\xi_{m}(t)\,\xi_{n}(t')} = \Gamma\,\delta_{m,n}\,\delta(t-t') $.
Here, $ \Gamma = \tilde{\Gamma}/\omega_{eg} $ is a dimensionless dephasing rate. In microscopic derivations of the HSR model as the high-temperature limit of a bosonic bath, $\Gamma$ scales linearly with $T$; in the present work, we treat $\Gamma$ as a phenomenological parameter controlling dephasing strength.

\begin{figure}[h]
    \centering
    \includegraphics{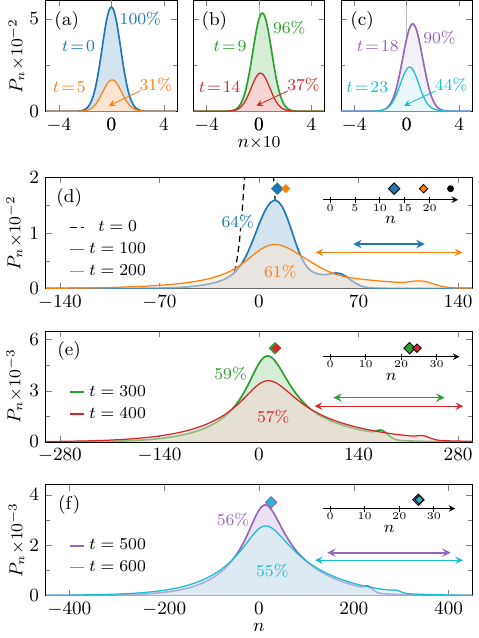} 
    \caption{\textbf{Noise-driven stages of real-space
    wave-packet dynamics.}
    (a-c)\;Short-time Rabi-oscillation regime. (d-f)\;Emitter-population
    distributions $P_n$ at later times. Filled distributions and time labels
    share the same color in each panel. Diamond markers denote the
    center-of-mass (CM) position, and double-headed arrows denote the spatial
    width of $P_n$ (proportional to the standard deviation). The black dot in
    the upper-right corner of panel (d) marks the CM position in the absence of
    noise, highlighting the slowdown of CM propagation induced by dephasing.
    The instantaneous total excitonic population is given as a percentage.
    Parameters: $\uprho=1$, $\omega_{c}=0.4$, $g=0.3$, $\Gamma=0.03$, $p=0.5$,
    $w=10$.}
    \label{fig:FIG6}
    \end{figure}

One way to capture such dynamics is by solving the time-dependent
Schr\"odinger equation for many noise realizations and then averaging over
them. However, for white noise (i.e., delta-correlated in time), it is
possible to perform an exact noise averaging analytically, resulting in a
master equation for the $ 2N \times 2N $ density matrix $ \hat{\rho} $ of
the exciton--photon system \cite{KenkreReineker1982}:
\begin{equation}
\frac{\partial \hat{\rho}}{\partial t} = -i\bigl[\hat{H}, \hat{\rho}\bigr]
-\frac{\Gamma}{2}\sum_{n=1}^{N}
\Bigl[\ket{b_{n}}\bra{b_{n}},
\bigl[\ket{b_{n}}\bra{b_{n}}, \hat{\rho}\bigr]\Bigr]. \label{eq:HSR-master-equation}
\end{equation}

Because the white-noise averaging effectively replaces the time-dependent
fluctuations $ \xi_{n}(t) $ with a static dephasing term proportional to 
$\Gamma$, the resulting master equation appears as time-independent 
(beyond the usual Hamiltonian evolution). Physically, the term $\propto\Gamma$ 
diminishes both the coherences within the excitonic subspace and the
light-matter coherences (i.e., off-diagonal elements in $\hat{\rho}$),
reflecting the loss of phase coherence induced by the noise. A complete
derivation of this result can be found in \cite{KenkreReineker1982}.\\

In the presence of noise, the dynamics depend on the number 
density,  $\uprho$.
As we discuss later, noise homogenizes the initial $k$-space distribution, and the 
scaling of the initial state variables becomes irrelevant on the long-time scale.
Thus, in this section, we fix the system parameters:
$\omega_{eg}=2.1\,\mathrm{eV}$, $\ell=0.590\mu \mathrm{m}$, $L=590\mu \mathrm{m}$. 
\add[IT:]{For simplicity, we set} the number of photonic modes equals the number of emitters,
 $N=1001$, such that $\uprho=1$.

Figure\;\ref{fig:FIG6} shows snapshots of emitters' probability
distributions in the presence of noise, obtained by direct numerical
solution of Eq.\;\eqref{eq:HSR-master-equation}. The figure suggests that the 
wave packet evolves through a sequence of qualitatively different 
relaxation stages across several time scales.
On the short time scale presented in Fig.\;\ref{fig:FIG6}(a-c),
the CM and the wave packet width remain almost unchanged, while the
the distribution's peak oscillates due to the coherent population
exchange between the light and matter subsystems
(Rabi oscillations). 
After the initial phase of rapid exchange, the population relaxes
slowly. Relaxation lasts until an equal population partition between
the subsystems is reached.

Although the initial center-of-mass (CM) velocity of the wave packet matches
that of the noise-free case, noise induces a gradual slowdown, leading to
a complete halt at $t \approx 200$. In contrast, the wave packet width
continues to grow even after the CM motion stops 
Thus, the stages of the matter wave packet evolution are (from
the shortest to longest): (a)\;underdamped Rabi oscillations, (b)\;damping
of CM motion, (c)\;population relaxation, (d)\;transition from 
ballistic to diffusive spreading.

\begin{figure}
\begin{centering}
\includegraphics{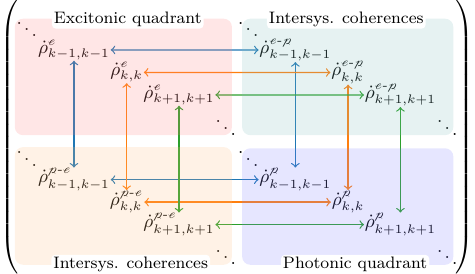}
\end{centering}
\caption{\textbf{Bloch-space density-matrix structure under
dephasing.} Illustration of the $k$-space density matrix.
Diagonal elements within the four quadrants (elements with $q=0$) and the
couplings between them are shown. The red shaded block denotes the excitonic
quadrant, the blue shaded block denotes the photonic quadrant, and the teal and
orange shaded blocks denote the two inter-system coherence quadrants.
Double-headed arrow boxes in matching colors indicate representative couplings
for the $k-1$, $k$, and $k+1$ mode pairs.}
\label{fig:FIG7}
\end{figure}

For further analysis, we utilize the spatial invariance of the emitters'
lattice and rewrite the master equation in terms of the Bloch states. 
This results in a reduced set of equations for the $k$-space density matrix elements
\begin{subequations}
\label{eq:Reduced-rho-equations} 
\begin{align}
\dot{\rho}_{k,k+q}^{\e}\!=\!-2ig & \mathrm{Im}[\rho_{k,k+q}^{\pe}]\!-\!\Gamma\rho_{k,k+q}^{\e}\!+\!\frac{\Gamma}{N}\sum_{p}\rho_{p,p+q}^{\e},
\label{eq:dens-mat-eqs-1}\\
\dot{\rho}_{k,k+q}^{\p}=-i & (\omega_k-\omega_{k+q})\rho_{k,k+q}^{\p}-2ig\mathrm{Im}[\rho_{k,k+q}^{\ep}],
\label{eq:dens-mat-eqs-2}\\
\dot{\rho}_{k,k+q}^{\pe}=-i & (\omega_{k}-\omega_\mathrm{eg}-i\Gamma/2)\rho_{k,k+q}^{\pe}\nonumber \\
-ig & (\rho_{k,k+q}^{\e}-\rho_{k,k+q}^{\p}),
\label{eq:dens-mat-eqs-3}\\
\dot{\rho}_{k,k+q}^{\ep}=-i & (\omega_\mathrm{eg}-\omega_{k+q}-i\Gamma/2)\rho_{k,k+q}^{\ep} \nonumber \\
-ig & (\rho_{k,k+q}^{\p}-\rho_{k,k+q}^{\e}),
\label{eq:dens-mat-eqs-4}
\end{align}
\end{subequations}
where $\rho_{k,k+q}^{\e}$ is an element in the first quadrant (excitonic
quadrant), $\rho_{k,k+q}^{\p}$ is an element in the fourth quadrant
(cavity or photonic quadrant), and $\rho_{k,k+q}^{\pe}$, $\rho_{k,k+q}^{\ep}$
are elements in the second and third, i.e., off-diagonal quadrants
of the density matrix. Setting $q=0$
shows that the populations of $k$-th exciton and $k$-th cavity
modes are coupled through light-matter coherence. Fig.\;\ref{fig:FIG7}
illustrates the $k$-space density matrix and shows some of its elements 
with $q=0$.

\begin{figure*}
\begin{centering}
\includegraphics{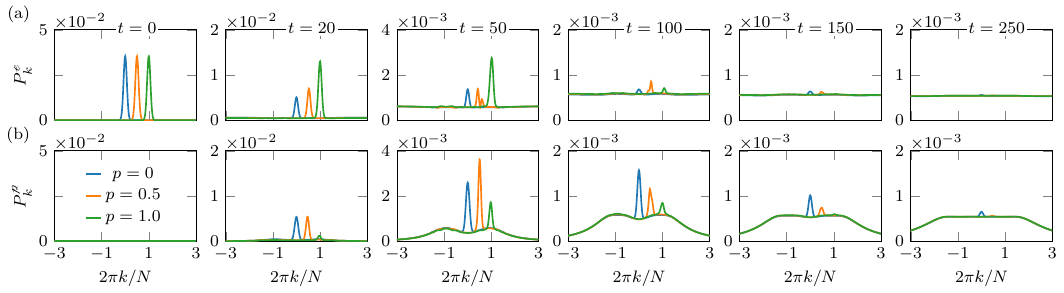} 
\end{centering}
\caption{\textbf{Population redistribution in momentum space
during relaxation.} 
Row (a) shows the excitonic distributions $P_k^{\e}$, and row (b) shows the
photonic distributions $P_k^{\p}$, at selected times. As time increases, the
excitonic distribution in row (a) becomes symmetric about $k=0$, which
coincides with the stopping of the CM motion. In each panel, the three colored
curves correspond to initial momenta $p=0$, $p=0.5$, and $p=1.0$. The columns
correspond to $t=0,20,50,100,150,250$ from left to right. Parameters:
$\uprho=1$, $\omega_{c}=0.4$, $g=0.3$, $\Gamma=0.1$, $w=10$.}
\label{fig:FIG8}
\end{figure*}

The effect of noise is two-fold. First, it damps the light-matter
coherences at rate $\Gamma/2$. Secondly, the last two terms in Eq.\;\eqref{eq:dens-mat-eqs-1}
tend to homogenize the population among the exciton modes.
The interplay between $g$ and $\Gamma$ controls the hierarchy of dynamical stages: $g$ sets the short-time coherent light--matter exchange, while $\Gamma$ controls decoherence and mode homogenization. Because the system comprises many detuned mode pairs, the effective long-time rates extracted from observables can be much smaller than $\Gamma$, even though $\Gamma$ is the only explicit dissipative scale.
Figure\;\ref{fig:FIG8} shows snapshots of excitonic ($P_{k}^{\e}$)
and photonic $(P_{k}^{\p})$ distributions. On the short time scale,
the distributions remain at the initial positions defined by the initial momentum, $p$, and oscillate due to the 
exchange of the population between the light and matter subsystems
(Rabi oscillations). With time, due to the noise-induced homogenization,
baselines are built in the excitonic distributions. Simultaneously,
all photonic modes become gradually populated. Later,
at $t\approx100-200$, the excitonic distribution becomes
symmetric relative to $k=0$, which explains the termination
of CM motion, as the average momentum vanishes [see Fig.\;\ref{fig:FIG6}(e,f)].
The population relaxation continues until both $P_{k}^{\e}$
and $P_{k}^{\p}$ become uniform.
We note that this long-time $k$-space homogenization is a feature of our minimal TLS + Markovian dephasing model. More microscopic treatments with additional internal structure (e.g., vibronic couplings), disorder, or non-Markovian relaxation pathways can instead yield momentum-space bottleneck accumulation/localization. Such effects are therefore outside the scope of the present model and may explain differences with more realistic simulations~\cite{AroeiraRibeiro2023, Krupp2025nature_comm}.

\subsection{Population relaxation: thermalization process}

\begin{figure}
\begin{centering}
\includegraphics{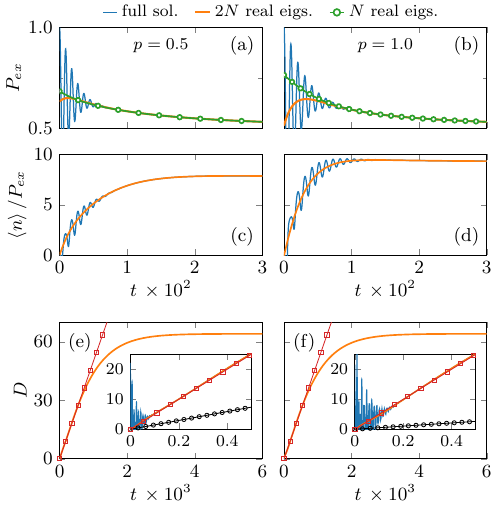} 
\end{centering}
\caption{\textbf{Population, center-of-mass motion, and
diffusivity under noise.}
(a,b)\;Total excitonic population. (c,d)\;Position of the matter wave packet
CM. (e,f)\;Diffusivity of the matter wave packet. Panels (a,b): solid colored
lines show the full numerical solution, thicker colored lines show
reconstruction from $2N$ real eigenvalues, and open-circle marker lines show
reconstruction from $N$ real eigenvalues. Panels (c,d): colored lines show CM
dynamics for $p=0.5$ and $p=1.0$. 
Panels (e,f): Time-dependent diffusivity $D(t)$. The main panels show the noisy dynamics (solid lines) together with the ballistic-regime linear estimate (open red squares), $[a(1-e^{-1}) - b/e]t$ from Eq.~\eqref{eq:D-linear-approx}. The fitted slope is $0.05$ and is approximately independent of $\Gamma$ and $p$. \textit{The isolated reference ($\Gamma = 0$, open black circles) is shown only in the insets}, where short-time behavior is compared. Insets in (e,f): black circles and blue lines denote the isolated and noisy dynamics, respectively, while red squares indicate the linear estimate. Parameters: $\uprho=1$, $\omega_c=0.4$,
$g=0.3$, $w=10$, $\Gamma=0.1$.}
\label{fig:FIG9}
\end{figure}

Figure~\ref{fig:FIG9}(a,b) shows the evolution of the total excitonic population in the presence of noise. Unlike in the isolated (noise-free) case, the Rabi oscillations decay on a timescale of $2/\Gamma$, regardless of the initial momentum $p$.
The
population slowly relaxes towards the equilibrium value of $1/2$
on the longer time scale. This behavior is expected as the thermalization
is independent of the initial state.
In particular, in the long-time limit the dephasing suppresses coherences and, together with the noise-induced mode homogenization in Eq.~\eqref{eq:dens-mat-eqs-1}, drives the system toward an effectively infinite-temperature incoherent mixture in the single-excitation exciton--photon subspace. In this limit the excitonic and photonic sectors are equipartitioned, yielding $P_{ex}(t\to\infty)=1/2$.
The relaxation time scales can be roughly estimated by neglecting
the homogenizing terms in Eq.\;\eqref{eq:dens-mat-eqs-1}, which
may be justified when the excitonic population distribution is
approximately uniform. Then, the system reduces to a collection 
of independent TLS where only the $k$-th exciton is coupled to 
stochastic noise i.e. stochastic $k$-dependent JC model. Population dynamics in such TLS has an 
oscillating contribution that decays at a rate $\sim \Gamma/2$, 
and decaying  at rate $s_{k}$ contribution, where
 \add{(see Supplementary Note 2 for details)} 
\begin{equation}
s_{k}\approx\frac{2g^{2}\Gamma}{4g^{2}+\delta_{k}^{2}},\quad\delta_{k}=\omega_{k}-\omega_{\mathrm{eg}}.
\label{eq:s_k}
\end{equation}
With the homogenizing terms, all the matter modes become populated
with time. The detuning of most modes from the corresponding
cavity modes gives rise to the slow relaxation seen in Fig. \ref{fig:FIG9}(a,b).

\begin{figure}
\begin{centering}
\includegraphics{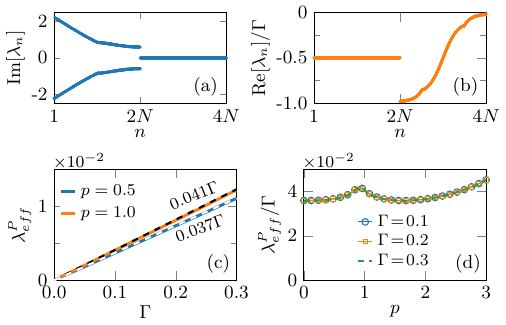} 
\end{centering}
\caption{\textbf{Spectral estimates of relaxation rates in the
reduced model.}
(a,b)\;Eigenvalues of the linear system in Eq.\;\eqref{eq:Reduced-rho-equations}.
Here, $q=0$, $\Gamma=0.1$. (c,d)\;Effective relaxation rate of the total
excitonic population, $P_{ex}$, see Eq.\;\eqref{eq:lambda-eff-P}. Panel (a):
blue point cloud gives $\mathrm{Im}[\lambda_n]$. Panel (b): orange point cloud
gives $\mathrm{Re}[\lambda_n]/\Gamma$. Panel (c): colored solid curves
correspond to $p=0.5$ and $p=1.0$; dashed guide lines indicate linear trends
$0.041\Gamma$ (black) and $0.037\Gamma$ (light dashed). Panel (d): open
circles, open squares, and dashed curves correspond to $\Gamma=0.1$, $0.2$,
and $0.3$, respectively. Note that in (d), $\lambda_{eff}^P$ is scaled by
$\Gamma$. Parameters: $\uprho=1$, $\omega_c=0.4$, $g=0.3$, $w=10$.}
\label{fig:FIG10}
\end{figure}

For quantitative time scale analysis, we consider the eigenvectors
and eigenvalues of the $4N\times4N$ non-Hermitian matrix defining
the linear system in Eq.\;\eqref{eq:Reduced-rho-equations}.
The total exciton population can be expressed as $P_{ex}=\sum_{n=1}^{4N}c_{n}\exp(\lambda_{n}t)$,
where $\lambda_{n}$ are the eigenvalues, and $c_{n}$ depend on the
eigenvectors and the initial state, and can be found by Moore--Penrose
pseudo-inverse. Figure\;\ref{fig:FIG10}(a,b) shows an example set
of eigenvalues, $\lambda_{n}$. All the eigenvalues with non-zero
imaginary parts have nearly constant real parts, $-\Gamma/2$. This
implies that the oscillations decay at a rate of $\Gamma/2$ as seen
in Fig.\;\ref{fig:FIG9}(a,b). Considering only the real eigenvalues
(with zero imaginary part) results in the smooth orange curves.\\

To determine the relaxation time scale, we focus on the $N$ smallest
real-valued eigenvalues ($-\Gamma/2<\lambda_{n}\le0$). Only those
eigenvalues contribute to the marked green curves in Fig.\;\ref{fig:FIG9}(a,b).
We found numerically that these eigenvalues have linear scaling with
$\Gamma$ (for $\Gamma\leq0.3$) \add{(see Supplementary Note 3 and Fig.~S1)}.

Processes described by a sum of decaying exponentials are encountered
e.g., in studies of time-resolved luminescence spectroscopy of molecular,
macromolecular, supramolecular, and nanosystems \cite{BerberanSantos2005}.
A stretched exponential (Kohlrausch) function is often used to model
and quantify such processes. Excellent fits (the fit quality
was assessed from the residual sum of squares) are obtained using the
model $a\exp(-\alpha t^{\beta})+1/2$,  and $\beta$ is found to be
close to $2/3$ \add{(see Supplementary Note 4 and Fig.~S4)}.

For simplicity and physical clarity, however, here we estimate the
effective decay rate of the population, $\lambda_{eff}^{P}$ by 
\begin{equation}
[\lambda_{eff}^{P}]^{-1}=\frac{\int_{0}^{\infty}t[f(t)-1/2]\,dt}{\int_{0}^{\infty}[f(t)-1/2]\,dt},
\label{eq:lambda-eff-P}
\end{equation}
where $f(t)$ is the population signal. When $f(t)$ taken to be the
smoothed out signal, $\lambda_{eff}^{P}$ is proportional to $\Gamma$
as shown in Fig.\;\ref{fig:FIG10}(c). This is consistent with the
linear $\Gamma$ dependence of the smallest real-valued eigenvalues.
When $f(t)$ is taken to be the full signal including the oscillations,
$\lambda_{eff}^{P}(\Gamma)$ shows slight deviation from the linear
$\Gamma$-dependence \add{(see Supplementary Note 5 and Fig.~S5)}.
These findings suggest that population relaxation is slower than Rabi
oscillation decay. Averages of $N/2$ smallest eigenvalues and averages
of the corresponding TLS rates, $s_{k}$ are comparable and can be
used as rough estimates of $\lambda_{eff}^{P}$ \add{(see Supplementary Note 3, Fig.~S2 for the eigenvalue-average comparison, and Fig.~S3 for the corresponding CM dynamics)}.
Figure\;\ref{fig:FIG10}(d) shows that $\lambda_{eff}^{P}$ has a
relatively weak dependence on the initial momentum $p$.
\subsection{CM motion of the matter distribution}

\begin{figure}
\begin{centering}
\includegraphics{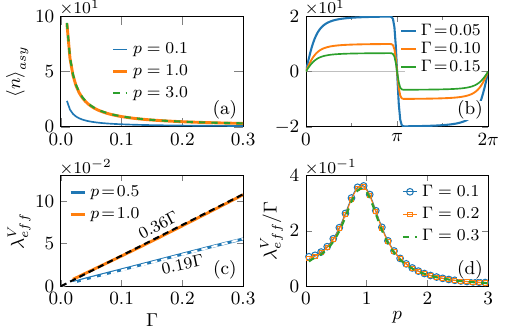} 
\end{centering}
\caption{\textbf{Effective center-of-mass stopping distance and
damping rate.}
Asymptotic position of the matter CM distribution as a function of $\Gamma$ (a)
and initial momentum, $p$ (b). The periodicity in (b) stems from the $k$-space
periodicity. (c,d)\;Effective decay rate of the CM velocity in
Eq.\;\eqref{eq:lambda-eff-V}. In (a,b), three line styles denote different
initial conditions (solid, thick solid, and dashed as labeled in-panel). In
(c), colored solid lines correspond to $p=0.5$ and $p=1.0$, while black and
light dashed lines are linear guides. In (d), open circles, open squares, and
dashed lines correspond to $\Gamma=0.1$, $0.2$, and $0.3$, respectively.
Parameters: $\uprho=1$, $\omega_{c}=0.4$, $g=0.3$, $w=10$.}
\label{fig:FIG11}
\end{figure}

Figure\;\ref{fig:FIG9}(c,d) shows two examples of matter distribution
CM dynamics. \add{Supplementary Note 6} outlines our approach
to evaluating $\braket{n(t)}$ using Eq.\;\eqref{eq:Reduced-rho-equations}.
The smooth orange curves were obtained by including only the real-valued
eigenvalues. The noise slows down and eventually stops the CM motion.
Qualitatively, this is similar to propagation on a simple 1D quantum
lattice (in the HSR model). In that case, the CM position is given
by \cite{Tutunnikov2023} 
\begin{equation}
\braket{n}=\frac{1-e^{-\Gamma t}}{\Gamma}\int_{-\pi}^{\pi}\varepsilon'_{k}P_{k}(0)\,dk,
\label{eq:<n>-lattice}
\end{equation}
where $\varepsilon'_{k}\equiv d_{k}(\varepsilon_{k})$, and $\varepsilon_{k}$
is the lattice dispersion relation. The formula shows that the asymptotic
position is proportional to $\Gamma^{-1}$ for a fixed initial momentum.
For comparison, Fig.\;\ref{fig:FIG11}(a,b) shows the asymptotic
values of the CM 
\begin{equation}
\braket{n}_{asy}\equiv\frac{\braket{n(t\rightarrow\infty)}}{P(t\rightarrow\infty)}=2\braket{n(t\rightarrow\infty)},
\end{equation}
where we used the fact that $P(t\rightarrow\infty)=1/2$. The $k$-space
periodicity is reflected in the periodicity of $\braket{n}_{asy}$
plotted as a function of $p$.

Despite the similarities with the simple lattice case outside a cavity,
cavity system exhibits very different behavior quantitatively. The CM
velocity decay rate here is significantly lower than $\Gamma$.
To estimate this rate, we use 
\begin{equation}
[\lambda_{eff}^{V}]^{-1}=\frac{\int_{0}^{\infty}t|f(t)|\,dt}{\int_{0}^{\infty}|f(t)|\,dt},
\label{eq:lambda-eff-V}
\end{equation}
where $f(t)$ is the velocity derived from the smooth orange curves
in Fig.\;\ref{fig:FIG9}(c,d). Figure\;\ref{fig:FIG11}(c,d) shows
$\lambda_{eff}^{V}$ as a function of $\Gamma$ and $p$. Close to resonance
($p\approx\omega_{\mathrm{eg}}$), the decay rate of the CM velocity 
is an order of magnitude larger than the population relaxation.

Qualitatively, the difference in population relaxation and CM velocity
decay rates can be seen through the population evolution of matter and cavity
modes. Figure\;\ref{fig:FIG8} shows that by the time $t\approx 150$ the excitonic distribution becomes nearly
symmetric and uniform. The symmetry implies vanishing average momentum,
which is consistent with Fig.\;\ref{fig:FIG9}(c,d) where the CM
indeed stops after $t\approx200$. In contrast, the cavity
distribution is not yet homogeneous, and the population exchange between
the subsystems continues.

\subsection{Diffusivity: ballistic-diffusive transition}

Next, we consider the rate of the matter distribution spreading, 
i.e., the diffusivity [compare with Eq.\;\eqref{eq:W}] 
\begin{equation}
2D(t)\equiv\frac{d}{dt}\left[\frac{\braket{n(t)^{2}}}{P_{ex}(t)}-\frac{\braket{n(t)}^{2}}{P_{ex}^{2}(t)}\right].
\label{eq:diffusivity-definition}
\end{equation}
\add{Supplementary Note 6} outlines our approach to evaluating
$\braket{n^{2}(t)}$ using Eqs.\;\eqref{eq:Reduced-rho-equations}.
Figure\;\ref{fig:FIG9}(e,f) shows two examples of the time-dependent
diffusivity that undergoes a transition from ballistic $(D\propto t)$
to diffusive $(D\approx\mathrm{const.})$ behavior. The transition
is associated with noise-induced dephasing and the homogenizing terms
in Eq.\;\eqref{eq:dens-mat-eqs-1}.
We emphasize that $D(t)$ quantifies the broadening of the excitonic probability distribution (units of length$^{2}$/time) and should not be conflated with the polariton group velocity $v_g=\partial\varepsilon_{k_\pm}/\partial k$ (units of length/time). In our model, dephasing can increase $D(t)$ for extended initial wave packets by suppressing interference between momentum components, even though this does not imply an increase of $v_g$~\cite{Ying2025nature_comm}.

In the case of 1D quantum lattice with HSR noise, the diffusivity
of a wave packet with stationary CM is given by 
\begin{subequations}
\begin{align}
D(t) & =C_{1}\frac{1-e^{-\Gamma t}}{\Gamma}+C_{2}e^{-\Gamma t}t,
\label{eq:D-in-HSR-lattice}\\
C_{1} & =\int_{-\pi}^{\pi}\frac{(\varepsilon'_{k})^{2}}{2\pi}\,dk,\\
C_{2} & =\int_{-\pi}^{\pi}(\varepsilon'_{k})^{2}\left[P_{k}(0)-\frac{1}{2\pi}\right]\,dk.
\end{align}
\end{subequations}

\begin{figure}
\begin{centering}
\includegraphics{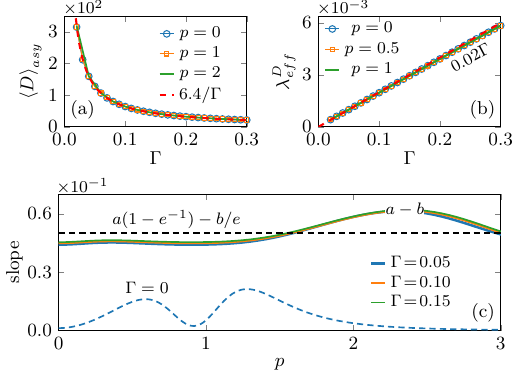} 
\end{centering}
\caption{\textbf{Noise-dependent diffusion constant and
crossover timescale.}
(a)\;Diffusion constant. (b)\;Effective ballistic-diffusive transition rate.
(c)\;Approximate slope of $D(t)$ in the presence of noise during the ballistic
stage, $a(1-e^{-1})-b/e$, $(a-b)$ [see Eqs.\;\eqref{eq:D-model} and
\eqref{eq:D-linear-approx}] vs.\ the slope in the isolated system
($\Gamma=0$). In (a,b), open circles, open squares, and solid curves denote
$p=0$, $0.5$, and $1.0$, respectively; red dashed guides indicate $6.4/\Gamma$
in (a) and $0.02\Gamma$ in (b). In (c), solid curves correspond to finite-noise
cases ($\Gamma=0.05,0.10,0.15$), the black horizontal dashed line denotes
$a(1-e^{-1})-b/e$, and the black sloped dashed reference denotes the
isolated-case slope ($\Gamma=0$). Parameters: $\uprho = 1$, $\omega_{c}=0.4$,
$g=0.3$, $w=10$.}
\label{fig:FIG9_12}
\end{figure}

Figure\;\ref{fig:FIG9_12}(a) presents the diffusion constant, $D_{asy}\equiv D(t\rightarrow\infty)\approx a/\lambda_{eff}$ \add{(detailed derivation of the above result can be found in Supplementary Note 7 and Ref.~\cite{Tutunnikov2023})}.
For the considered values of $\Gamma$ ($0.01 \leq \Gamma \leq 0.3$),
$D_{asy}$ scales as $\Gamma^{-1}$ independent of the initial state.
For comparison, the diffusion constant in the HSR model, $J$ is given by $D_{lattice}(t\rightarrow\infty)=2J^{2}/\Gamma$.

To study the time dependence of diffusivity, we use the model suggested
by Eq.\;\eqref{eq:D-in-HSR-lattice} 
\begin{equation}
f(t)=a\frac{1-e^{-\lambda_{eff}^{D}t}}{\lambda_{eff}^{D}}-be^{-\lambda_{eff}^{D}t}t.
\label{eq:D-model}
\end{equation}
This model has three free parameters: $a$, $b$, and $\lambda_{eff}^{D}$,
and provides excellent fits for the system parameters considered.
Figure\;\ref{fig:FIG9_12}(b) shows $\lambda_{eff}^{D}$ as a function
of $\Gamma$. While the ballistic-diffusion transition rate in the
lattice is $\Gamma^{-1}$, the rate here is significantly slower,
$\lambda_{eff}^{D}=0.02\Gamma$. Like the HSR model,
the rate is independent of the initial momentum.

Importantly, the transient diffusivity in the ballistic regime is
significantly enhanced compared to the diffusivity in the isolated
system. Figure\;\ref{fig:FIG9}(e,f) compares the diffusivity in
the isolated system (black circles) vs the linear approximation denoted
by red squares, 
\begin{equation}
\lambda_{eff}^{D}f(1/\lambda_{eff}^{D})t=[a(1-e^{-1})-b/e]t.
\label{eq:D-linear-approx}
\end{equation}
The slope, $a(1-e^{-1})-b/e \approx 0.05$ (for $\uprho=1$, $\omega_{c}=0.4$, $g=0.3$) is independent of $p$ and $\Gamma$
for the considered values of $\Gamma\geq0.02$. Noise-induced
transient diffusivity enhancement has been studied in the HSR model \cite{Tutunnikov2023}. In our multimode cavity
system, the duration of the enhanced expansion is significantly
extended due to the slow ballistic-diffusion transition rate. Figure
\;\ref{fig:FIG9_12}(c) compares the approximate slope of $D(t)$ in
the ballistic regime, $a(1-e^{-1})-b/e$ vs the slopes in the isolated
system (dashed curve). We added the initial slope of $D(t)$ and $(a-b)$
for comparison.
\subsection{Dependence on the initial excitation:  Polariton vs Excitonic Initial States} \label{sec:pol_vs_excSystem-definition}
In this section, we examine the influence of the initial state on the wave-packet dynamics in the presence of dephasing noise. In addition to the excitonic initial condition studied earlier, we now consider wave packets initialized directly in the lower-polariton (LP) or upper-polariton (UP) branch\rep[ER-LPUP-DEF]{}{.For each momentum $k$, the single-excitation block of the multimode TC Hamiltonian admits two eigenstates, $|v_k^{\pm}\rangle$, with eigenvalues $\varepsilon_{k}^{\pm}$ given by the polariton dispersion relations (cf.~Fig.~\ref{fig:FIG1}(b)). These polariton eigenstates are coherent superpositions of the excitonic mode $|b_k\rangle$ and the photonic mode $|a_k\rangle$ at the same momentum, with mixing angle set by the detuning $\delta_k=\omega_k-\omega_{\mathrm{eg}}$ and the coupling strength $g$. The corresponding polariton wave packet is then}
\begin{equation}
|\psi_{\pm}(t)\rangle = \sum_{k} \psi_k(0)\, e^{-i \varepsilon_k^{\pm} t}\, |v_k^{\pm}\rangle,
\end{equation}
where $\pm$ \rep[ER-LPUP-DEF]{}{select the upper- or lower-polariton branch, and $\psi_k(0)$ is the (normalized, $\sum_k |\psi_k(0)|^2=1$) initial momentum-space distribution, taken here to be the same Gaussian as in the excitonic case to allow direct comparison}.
\begin{figure}
\begin{centering}
\includegraphics[]{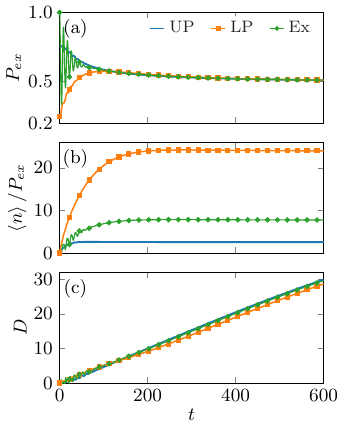}
\end{centering}
\caption{\textbf{Dependence on the initial excitation.} Dynamics of wave packets initialized in different excitation manifolds in the presence of dephasing noise. 
(a) Total excitonic population $P(t)$, (b) normalized center-of-mass position $\langle n(t)\rangle / P(t)$, and 
(c) time-dependent diffusivity $D(t)$ for wave packets initialized in the upper polariton (UP), lower polariton (LP), 
and excitonic (Ex) states. While the excitonic initial condition exhibits transient Rabi oscillations at short times, all three initializations 
show nearly identical relaxation dynamics at longer times due to noise-induced dephasing. In contrast, the 
center-of-mass motion remains strongly dependent on the initial state, reflecting the distinct group velocities 
associated with the polariton branches. Parameters: $\rho = 1$, $\omega_c = 0.4$, $g = 0.3$, $w = 10$, $\Gamma = 0.1$, and $p = 0.5$.
}
\label{fig:FIG9_ep}
\end{figure}
The dynamical results are shown in Fig.~\ref{fig:FIG9_ep} for a representative noise strength $\Gamma = 0.1$ and $p = 0.5$.
All other parameters are the same as before.  
All three moments [i.e., population in Fig.~\ref{fig:FIG9_ep}(a), average position in Fig.~\ref{fig:FIG9_ep}(b),
and diffusivity in Fig.~\ref{fig:FIG9_ep}(c)] show essentially the same relaxation time scales,
except for the transient Rabi oscillation present in the excitonic initial state.
The damping of the Rabi oscillation occurs on the dephasing time and represents the shortest time scale.
The subsequent three timescales, i.e., slowdown of the center of mass motion, population relaxation, 
and ballistic-to-diffusive transitions, are independent of the initial condition.  

Further, beyond the transient oscillations, the time evolutions of population and diffusivity of the three initial conditions become nearly identical, as the dephasing noise suppresses the non-Makovian effect in polariton dynamics.  The exception is the normalized center-of-mass motion, shown in Fig.~\ref{fig:FIG9_ep}(b), which retains a strong dependence on the initial state. In particular, LP, UP, and exciton-initialized wave packets propagate with distinct velocities, reflecting the underlying polariton dispersion and the branch-selective nature of the excitation. 

Overall, while the ballistic transport properties (such as Rabi oscillations and center-of-mass propagation) remain sensitive to the initial state, the long-time population dynamics and diffusive behavior are largely universal in the presence of noise. Thus, our early predictions based on the excitonic initial condition can be observed regardless of the initial excitation condition. 

\section{Discussion\label{sec:Conclusions}}

The presented theoretical analysis characterizes the essential
features of polariton wave packet dynamics in an effectively 1D multi-mode microcavity with stochastic noise acting on the emitters.

Noise acting on the emitters induces dephasing in the system. It causes significant
changes in the dynamics of the emitter's population distribution, which proceeds through
a sequence of relaxation stages across vastly different time scales. These stages include Rabi oscillations, CM velocity damping, population relaxation, and the transition from the ballistic to the diffusive regime of distribution expansion. 
Though the ballistic behavior depends on the initial preparation, the diffusive behavior becomes largely independent of the initial state, as discussed in Sec.~\ref{sec:pol_vs_excSystem-definition}.
The time scales of these stages establish a universal hierarchy,
$\Gamma^{-1} < (\lambda_{\text{eff}}^V)^{-1} < (\lambda_{\text{eff}}^P)^{-1}
< (\lambda_{\text{eff}}^D)^{-1}$.
Here, $\Gamma$ is the white noise's intensity. The initial stage, characterized by the
rapid decay of Rabi oscillations, occurs on the time scale of $\Gamma^{-1}$.
$\lambda_{\text{eff}}^V$ is the rate of CM velocity damping,
$\lambda_{\text{eff}}^P$ is the population relaxation rate, and
$\lambda_{\text{eff}}^D$ is the rate of ballistic--diffusive transition.
Importantly, $\lambda_{\text{eff}}^V$, $\lambda_{\text{eff}}^P$, and
$\lambda_{\text{eff}}^D$ can be smaller than $\Gamma$ by several orders of magnitude.
This contrasts sharply with wave packet dynamics in simpler 1D lattices without cavities
(including direct inter-emitter interactions) i.e., the HSR model, where all time-dependent observables have a
single characteristic time scale, $\Gamma^{-1}$.
Another consequence of noise is that the initial ballistic expansion
persists for several orders of magnitude longer than the microscopic dephasing time
$\Gamma^{-1}$. Moreover, the rate of expansion is substantially higher in the presence of noise than in the noise-free case. 

Finally, we assess the role of the initial excitation and find a clear separation between transient and long-time behavior. While short-time dynamics, such as Rabi oscillations and center-of-mass motion, depend on the initial state (exciton, LP, or UP), the long-time population dynamics and diffusivity become largely independent of the initial state due to dephasing. The only persistent distinction is in the center-of-mass motion, which reflects the branch-dependent group velocities. These results highlight that, although ballistic features are sensitive to the initial state, the long-time diffusive dynamics are universal.
Note that the model adopted here is a generalization of the HSR model \add{(see Supplementary Note 7)},
which assumes a classical stochastic noise, to a multimode cavity system
\cite{Chuang2016, Cao2009}. In other models, emitters' vibrational degrees 
of freedom are often treated classically or semi-classically~\cite{Sokolovskii2023, Luk2017}.
Such approaches may not fully capture the system coherences and become less 
reliable at low temperatures or under strong vibronic coupling \cite{CrespoOtero2018}. 
Coherences may also affect the thermalization process~\cite{svensden2024, schafer2019, xu2016}.
Thus, our model provides an exact quantum description of coherent exciton--photon dynamics in the presence of Markovian dephasing noise; it is not intended to capture low-temperature vibronic physics or finite-temperature energy relaxation, but it does capture the early-time ballistic expansion and the dephasing-induced crossover to diffusive-like spreading within this model.
These reservations also apply to other semi-classical simulations attempting 
to describe post-thermalization diffusion.
Polariton dispersion curves and the corresponding group velocities
are often used for interpreting experimental results focusing on the first two moments
of the population distribution in the matter subsystem
\cite{Xu2023, Jin2023, Pandya2022, Balasubrahmaniyam2023}.
However, microscopy measurements tracking the spatiotemporal profile of the wave packet
provide information beyond the first two moments.
In line with our analysis,
recent experiments have shown evidence of distribution splitting
\cite{Balasubrahmaniyam2023}.
The corresponding noise-free predictions
include momentum-dependent ballistic propagation and wave-packet splitting
within the same multimode cavity setting considered here.

Ballistic expansion at a fraction of the speed of light is expected in a noise-free and disorder-free polariton system.
Recent experiments demonstrated ballistic expansion,
despite disorder and dissipative effects
\cite{Balasubrahmaniyam2023, Jin2023, Xu2023, Allard2022, Pandya2022, Hou2020, schachenmayer2015}.
Our theoretical results are consistent with these experiments and predict
noise-enhanced ballistic expansion persisting on timescales several orders of
magnitude longer than the microscopic dephasing time $\Gamma^{-1}$.

A complementary way to benchmark the duration of the noise-enhanced regime is to estimate
the time $t^*$ at which the ballistic growth of the diffusivity overtakes the noisy plateau by equating
$D_{\Gamma=0}(t)\sim D_{\mathrm{asy}}$.
This criterion highlights the unusually long-lived ballistic regime induced by
dephasing in the multimode cavity system.
Observation of such a crossover in a lossy cavity would additionally require a photon
lifetime comparable to or longer than $t^*$.

The noise intensity $\Gamma$ is proportional to temperature, and all the time scales
related to dynamical observables were shown to scale with $\Gamma^{-1}$.
Both temperature and the initial state can be controlled experimentally, making our
theoretical predictions amenable to experimental verification.
However, long observation times may be necessary to confirm the predicted extended
ballistic spreading and the slow transition to a diffusive regime.
Inter-emitter interactions (such as F\"orster-like energy transfer
\cite{Anzola2020, jianshu2022jpcl, Du2018}), cavity losses, static disorder \cite{Engelhardt2023}, and other physical processes
may further influence ballistic expansion of the emitters' probability distribution.
In particular, radiative photon leakage from realistic microcavities (at an effective
rate $\kappa$) will compete with the long-time stages identified here.
Observation of population relaxation and the ballistic--diffusive crossover requires
$\kappa^{-1}$ to be comparable to or longer than the corresponding effective timescales;
otherwise the dynamics will be truncated, while the short-time coherent and early
dephasing regimes may remain accessible.

The effective master equation developed here allows for the straightforward inclusion
of some of these processes, particularly inter-emitter interactions and cavity loss,
paving the way for a more detailed description of specific experimental conditions.
A comprehensive description of realistic emitters, such as molecules, requires
accounting for vibrational degrees of freedom (i.e., vibronic coupling).
Future extensions will combine multimode cavity dynamics with vibronic structure
and non-Markovian environments to bridge the gap between idealized models and molecular
polaritonic experiments.

This paper is part of the series on the characterization of polariton wave-packet dynamics in a multimode cavity. The two other manuscripts explore the coherent-incoherence transition in lossy cavities 
\cite{Qutubu2025}
and free polariton wave-packet propagation with excitonic couplings, respectively. These projects will be followed by future studies of static disorder, vibronic coupling, phonon scatterings, inhomogeneous couplings etc.

\section*{Data Availability}
The data that support the findings of this study are available from the
corresponding author, J.C., upon request. The numerical datasets used for figure generation are stored in
the project data directories and institutional research storage maintained by
the authors.
\section*{Code Availability}
The custom code used to generate and analyze the results in this study is
available from the corresponding author, J.C., upon request. The code versions and the parameter sets
used for the reported calculations are documented in the manuscript, figure
captions, and Supplementary Information.
\section*{Author Contributions}
I.T., M.Q., H.R.S., and J.C. designed the research. I.T. and M.Q. performed the analytical and numerical calculations. 
All authors discussed the results and contributed to writing and revising the manuscript. J.C. guided the research project.
\section*{Competing interests}
The authors declare no competing interests.

\section*{Funding}
\add{J.C. and I.T. acknowledge support from the U.S. National Science Foundation (Grants No.~CHE-1800301 and No.~CHE-2324300) and the MIT Sloan Fund. I.T. and H.R.S. acknowledge support from the U.S. National Science Foundation through a grant for the Institute for Theoretical Atomic, Molecular, and Optical Physics (ITAMP) at Harvard University. M.Q. acknowledges financial and computational support from the Beijing Computational Science Research Center (CSRC).}

\begin{acknowledgments}
M.Q. extends thanks to G. Engelhardt, L. Premkumar, and K. E. Dorfman.
\end{acknowledgments}

\bibliography{bibliography}

\end{document}